\documentclass[sigconf]{acmart}
\AtBeginDocument{%
  \providecommand\BibTeX{{%
    \normalfont B\kern-0.5em{\scshape i\kern-0.25em b}\kern-0.8em\TeX}}}

\setcopyright{none}
\renewcommand\footnotetextcopyrightpermission[1]{}

\usepackage[utf8]{inputenc}
\usepackage[english]{babel}
\usepackage{hyperref}
\hypersetup{
    colorlinks=true,
    linkcolor=blue,
    filecolor=magenta,      
    urlcolor=cyan,
}

\newcommand{\mage}[1]{\textcolor{magenta}{#1}}

\begin{document}

\title{Thinking Through and Writing About Research Ethics Beyond ``Broader Impact''}

\vspace{10mm}
\author{Kate Sim}
\email{kate.sim@oii.ox.ac.uk}
\affiliation{%
  \institution{University of Oxford \\ Oxford Internet Institute}
  \city{Oxford}
  \country{UK}
}

\author{Andrew Brown}
\email{abrown@robots.ox.ac.uk}
\affiliation{%
  \institution{University of Oxford \\ Visual Geometry Group}
  \city{Oxford}
  \country{UK}
}

\author{Amelia Hassoun}
\email{amelia.hassoun@oii.ox.ac.uk}
\affiliation{%
  \institution{University of Oxford \\ Oxford Internet Institute}
  \city{Oxford}
  \country{UK}
}

\renewcommand{\shortauthors}{Sim, K. et al.}

\begin{abstract}
 In March 2021, we held the first installment of the tutorial on thinking through and writing about research ethics beyond ``Broader Impact'' in conjunction with the ACM Conference on Fairness, Accountability, and Transparency (FAccT '21). The goal of this tutorial was to offer a conceptual and practical starting point for engineers and social scientists interested in thinking more expansively, holistically, and critically about research ethics. This report provides an outline of the tutorial, and contains our ``lifecourse checklist''. This was presented as part of the tutorial, and provides a practical starting point for researchers when thinking about research ethics before a project's start. We provide this to the research community, with the hope that researchers use it when considering the ethics of their research. 
\end{abstract}

\begin{teaserfigure}
  \includegraphics[width=0.07\textwidth]{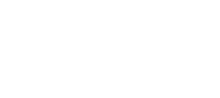}
  \Description{Enjoying the baseball game from the third-base
  seats. Ichiro Suzuki preparing to bat.}
  \label{fig:teaser}
\end{teaserfigure}


\maketitle

\section{Introduction}
The past year witnessed the computational field grapple with questions of responsibility and ethics of their research. Calls for papers in engineering conferences~\cite{ACL_20} include ``ethics'' and ``societal impact'' as topic areas. NeurIPS, for example, implemented the requirement that submissions include a ``broader impact statement''~\cite{NeurIPS_FAQ}. Conference organizers have also invited interdisciplinary reviewers to assess submissions’ discussion of social impact~\cite{NeurIPS_Program_Committee}. These efforts mark a long overdue, but exciting moment for the field as it asks difficult and important questions about its impact in the academic and social world. Reactions to these changes are instructive in evaluating whether and how they have been constructive in challenging researchers to critically examine their positionality, epistemology, and work’s impact during and throughout the research publication pipeline.

Barely a year into NeurIPS’ impact statement requirement, it appears its impact has been, at best, modest. Abuhamad and Rheault~\cite{abuhamad2020like} conducted a survey on this year’s NeurIPS authors’ experience with the broader impact requirement. Participants’ attitudes ranged from nonchalance and mild annoyance to outright hostility. Elsewhere, when papers do engage with “broader impact,” their engagement coheres around issues of bias and harm, according to Blodgett and colleagues~\cite{blodgett}. Surveying papers analyzing bias in natural language processing (NLP) submitted to the Association for Computer Linguistics, the authors found papers’ framing of bias to be “often vague, inconsistent, and lacking in normative reasoning,” have little to no engagement with literature outside of NLP, and preoccupied with computational fixes~\cite{blodgett}. 

These strike us as a missed opportunity. While ``bias'' and ``harms'' warrant sustained attention, simply identifying potential harmful impacts or quantifying bias constitutes a small facet of what critical voices in the field mean when they call for accountability in AI~\cite{automating_inequality,algo_oppress,race_after_tech}.  

This tutorial aims to offer a conceptual and practical primer for engineers interested in thinking more expansively, holistically, and critically about research ethics. We include issues of ``harm'' and ``bias'' as components of wider discussion about research ethics as it has been conceptualized and operationalized in the interpretivist traditions of humanities and the social sciences. In the interpretivist paradigm, ``doing ethical research is not as simple as following a set of rules''~\cite{willis_jost_nilakanta_2009}. Assessing the ethical impacts of research necessarily implicates the researcher’s standpoint~\cite{feminism_and_methodology,anthro} in the social world structured by power relations~\cite{srinivasan}, which, in turn, informs how they conducts their work~\cite{intersect}. In other words, what we see and how we see are inextricably shaped by who we are and what we do as~\cite{10.5555/782686.782694,10.1215/-12-2-1} knowledge workers~\cite{doi:10.1177/0001839212437519}. 

Meaningful engagement with research ethics must thus extend beyond diagnosing harms or anticipating carbon emissions and delve into this tripartite relationship. To do this, our goals are as follows: 

\begin{itemize}
\item Place interpretivist social scientific concepts of knowledge, power, and reflexivity in dialogue with engineering heuristics and practices for assessing research rigor, process, and impact
\item Provide concrete steps for thinking through research ethics during the publication pipeline 
\item Discuss why research ethics training is lacking in the field, and develop specific recommendations to program committees, senior researchers, etc. to incorporate into doctoral training
\end{itemize}

This is not to suggest that social scientific approaches to research ethics have the answers. In fact, much has been critiqued about the bureaucratization of research ethics, such as the Institutional Review Board (IRB) in the US~\cite{doi:10.1146/annurev.lawsocsci.093008.131454}. Rather, we see these approaches as an instructive model for identifying core values, setting norms, and operationalizing them into the research pipeline, from training to publication. 

Our goal is that the participants of our tutorial would walk away with the following: (1) actionable steps for discussing positionality and limitations in their writing; (2) a practical primer for mapping out ethical dimensions during the research process; and (3) specific interpretivist concepts like ``reflexivity'' to challenge their thinking about research purpose and process. 

The following report is organised as follows: in Section~\ref{overview} we provide an overview of the tutorial. In Section~\ref{checklist} we include the lifecourse checklist; a set of research ethics checkpoints that we provided with the tutorial. Slides from the tutorial can be found \href{https://docs.google.com/presentation/d/1fECOhSulRUuvK8L_4NA4AwuheF1HspbCTvqsoa6bx9U}{\mage{here}}, and a recording of the tutorial can be found \href{https://www.youtube.com/watch?v=M_iIN-N3y6w}{\mage{here}}.

\section{Tutorial Overview} \label{overview}
Due to COVID-19, and in line with FAccT 2021, the tutorial was held entirely virtually on Zoom. The tutorial spanned over 90 minutes, and had a participation of over 50 conference attendees throughout. Here, we provide an overview of the tutorial. The tutorial consisted of two parts:

The first part included a talk from the organisers on common misconceptions around research ethics and an imaginative exercise centred on the tutorial attendees' best and worst case scenarios for their own research projects. For the second part, the organisers outlined the lifecourse approach to research ethics. This was followed by the lifecourse group activity, and the following group discussion.

Inspired by the approach of Giele and Elder~\cite{elder1998methods}, the lifecourse approach to research ethics is centred on imagining each project as having a life course. It is up to the researcher to  then think about the conditions, assumptions, and aspirations that shape their research design in the three different stages of this lifecourse; before, during and after the project.

In the lifecourse group activity, participants were split into breakout rooms and each were given a machine learning case study to practice applying the lifecourse approach. Participants were given our lifecourse worksheet for guidance (provided in Section~\ref{checklist}). In the following group discussion, participants relayed their answers to the wider group, and discussed any difficulties/challenges they had as well as any interesting findings that arose from the approach.

\section{Lifecourse Checklist} \label{checklist}
Here, we provide the lifecourse checklist. This checklist was developed alongside the tutorial, and provides guidelines for applying the lifecourse approach when discussing research ethics before the start of a project. The checklist breaks down the lifecourse of a research project into three parts: before, during, and after. Some parts are not appropriate for certain research projects, and occasionally not enough information will be known about a project prior to its start to answer some questions. In these cases, we advise users to think about when in the project lifecourse you might be able to answer those questions.

\paragraph{\textbf{Before}}

\begin{itemize}
\item Informed Consent
\begin{itemize}
\item 1a) If your dataset contains humans, what efforts could you make to inform people that you want to use their data for your research, and seek their consent, even if using a large dataset? 
\item 1b) What expectations of privacy did people have when they provided the data you will be using for your study? How might your use of their data violate these expectations?
\item 1c) How could your research be using their data in ways they did not anticipate? 
\item 1d) What process will you use to discuss and document this, decide if your study violates their privacy, and come up with ways to mitigate this?

\end{itemize}

\item Opting Out
\begin{itemize}
\item 2a) If you can’t gain consent from each participant, how might you be able to still inform people that their data has been used, so they can opt-out?
\item 2b) Can people opt-out of your research? If so, when and how will you tell them how to do that? 
\item 2c) Who can they talk to if they have concerns, and how and when are you giving them this information? 
\item 2d) Can people slow down or  stop the research project entirely if they have enough concerns about its ethics? How?

\end{itemize}

\item Identifiablity 

\begin{itemize}
\item 3a) Can the identities of people in the dataset be deduced? 
\item 3b) Can you remove identifying information from the dataset? If so, how? 
\item 3c) Can it be re-identified?
\item 3d) How will de-identification alter the dataset’s representativeness? (\textit{e.g}. removing age from a dataset about Tik Tok users might reinforce the idea that only young people use the app).
\item 3e) How could someone represented in the dataset be potentially embarrassed or upset by it, even if information about them was removed?
\end{itemize}

\item Representativeness  

\begin{itemize}
\item 4a) Who (what groups/populations) does your dataset represent? How might your project frame overstate its representativeness? Who might it leave out or not represent? What harmful implications might this have?
\item 4b) If you need to make changes to the dataset or strip out identifying information (like race, gender, age, etc), how will the changes alter the original dataset’s representativeness? 
\item 4c) What assumptions does your dataset give you about social groups? (\textit{e.g.}, you may have gender parity in your dataset about profession, but “nurses” are female and “doctors” are male). 

\end{itemize}

\item Positionality 

\begin{itemize}

\item 5a) Who is conducting the study? What are the demographics of that team? How might those demographics affect how you frame your study and your blind spots? What biases and preconceived notions might they bring that will affect the way that the study is designed, carried out, and interpreted?)

\item 5b) What skills and expertise do you need to do this project ethically? What expertise do you have that allows you to evaluate these ethical questions fully? Who can you consult or bring on to help you fill any gaps you have?

\end{itemize}

\item Data Quality 

\begin{itemize}

\item 6a) Are you collecting more data than you need to answer your research question?

\item 6b) Why might collecting more data not always be better? 

\item 6c) Who could be harmed by over-collecting data? 

\item 6d) How might overcollection be harmful? 

\end{itemize}

\end{itemize}

\paragraph{\textbf{During}}

\begin{itemize}

\item Access To Study Information

\begin{itemize}

\item 7a) Where will you put information about the study so that it is clearly available to other academics and the public, so they can also reflect on its implications?

\end{itemize}

\item Data Storage \& Transfer

\begin{itemize}

\item 8a) Where are you putting the data you’re using? How is it safe? How is it protected? Who has access? For how long?

\item 8b) Will you transfer it anywhere, and if so how? What breaches could occur, and how will you mitigate these? 

\end{itemize}

\item Exposure To Risk

\begin{itemize}

\item 9a) For the people participating in this research (\textit{e.g.}, people data is collected from, communities, etc.): How could this research subject them to legal risk or disciplinary action? How could it make them or their data/activities visible to organisations (\textit{e.g.} states or corporations) in a way that would open them up to negative consequences?

\item 9a-1) How are people participating in the project safeguarded from these risks?

\item 9b) For the people working on this research (\textit{e.g.}, researchers, interns): What risk (mental, professional, or physical) could this research put those working on the project in?

\item 9b-2) How are the people working on this project safeguarded from these risks?

\end{itemize}

\item Exploitation

\begin{itemize}

\item 10a) How will you ensure that everyone working on the project is working fair hours in practice -- according to the labour laws of the university and the government -- and are not overworked? How will they be fairly compensated? 

\item 10b) How will you ensure that they are not being harassed or bullied? 

\item 10c) How can workers seek help if they are being exploited? How will you make everyone participating aware of these avenues?  

\item 10d) How are junior researchers working on the project (but who do not have a say in its direction/big picture decisions) being protected from the potential negative effects of the study, should the study draw negative critique?

\item 10e) What processes do you have in place to ensure that junior researchers can express concerns about the research without fearing retribution?

\end{itemize}

\item Ongoing Review

\begin{itemize}

\item 11a) What processes do you have to review the project during its duration to identify and fix anything that is not working as anticipated? 
\item 11b) How will you evaluate and address your method’s limitations or ethical issues that arise during the project as you apply them? 
\item 11c) Who will be invited to participate in these reviews? Who will not be invited to participate? 
 
\end{itemize}
\end{itemize}

\vspace{4mm}

\paragraph{\textbf{After}}

\begin{itemize}

\item Accessibility of Results
\begin{itemize}

\item 11a) Where will you publish your results? 
\item 11b) Who, or what groups, will have the most access to them? which groups will have less access to them?
\item 11c) How else will you disseminate your research? 
\item 11d) What steps will you take to make sure that your research’s limitations are clear to your different audiences? 

\end{itemize}

\item Usability
\begin{itemize}
\item 12a) Who is the intended audience of your results?

\item 12b)  How do you imagine them using the results or outcomes of your study? 

\end{itemize}

\item Immediate Impacts

\begin{itemize}
\item 13a) Who could the immediate results or findings of the project benefit? 
\item 13b) Who might be left out or not benefit?  Who might be negatively affected?
\item 13c) Who will make money from this research? How will you ensure this is as equitable as possible? 
\end{itemize}

\item Future Applications
\begin{itemize}
\item 14a) What are the potential positive future applications of your findings, and who might they positively affect? 

\item 14b) Which groups could be negatively affected by future applications of your research? How?  

\end{itemize}

\item Possible Misuses

\begin{itemize}
\item 15a) How could private organisations (like tech companies) misuse your research? How could public organisations (like governments or universities, in your country or elsewhere) misuse your research? 
\item 15b) How could the methods/datasets used or knowledge produced serve purposes other than ones intended?
\item  15c) How will you safeguard against misuse? 

\item 15d) What would your “words of caution” (i.e. datasheets) be to future users of the results? How will you inform/warn them against potential misuse?

\end{itemize}

\item Limitations

\begin{itemize}
\item 16a) Reflecting on your research, what are the actual possible benefits of the research? What are its limitations? Be honest!
\end{itemize}

\end{itemize}

\begin{acks}
Andrew Brown is funded by a EPSRC DTA Studentship. We would like to thank the members of the Visual Geometry Group in Oxford for their insightful discussions.
\end{acks}

\bibliographystyle{ACM-Reference-Format}
\bibliography{main}


\end{document}